\documentclass[twocolumn,prl,showpacs,preprintnumbers,amsmath,amssymb]{revtex4}

\usepackage{graphicx}
\usepackage{bm}
\usepackage{appendix}

\begin{document}

\title{Confinement of Skyrmion states in noncentrosymmetric magnets
}

\author{ A. A. Leonov,   A. N.\ Bogdanov, U. K. R\"o\ss ler}

\address{IFW Dresden, Postfach 270116, D-01171 Dresden, Germany}

\date{\today}

\begin{abstract}
{%
Skyrmionic states in noncentrosymmetric magnets 
with Lifshitz invariants are investigated 
within the phenomenological Dzyaloshinskii model.
The interaction between the chiral Skyrmions,
being repulsive in a broad temperature range, 
changes into attraction at high temperatures. 
This leads to a remarkable \textit{confinement} effect: 
near the ordering temperature Skyrmions 
exist only as bound states, and Skyrmion lattices 
are formed by an unusual instability-type nucleation transition.
Numerical investigations on two-dimensional 
models demonstrate the confinement 
and the occurrence of different Skyrmion 
lattice precursor states near the ordering transition 
that can become thermodynamically stable by 
anisotropy or longitudinal softness in cubic helimagnets.
We introduce a new fundamental parameter, 
\textit{confinement temperature} $T_p$ separating
the peculiar region with "confined" chiral modulations
from the main part of the phase diagram with regular 
helical and Skyrmion states. 
%
%
%
}
\end{abstract}


         
\maketitle

%

\vspace{5mm}

In certain noncentrosymmetric magnetic systems,
the asymmetric Dzyaloshinskii-Moriya (DM) exchange
results in chiral couplings that can stabilize
long-period, non-collinear modulations of 
the magnetization with a \textit{fixed} sense of
rotation \cite{Dz64,PRB02}.
These chiral couplings are phenomenologically described 
by Lifshitz invariants that destroy 
the homogeneity of ordered phases \cite{Dz64}.
%
%
In highly symmetric systems 
with Lifshitz invariants 
multiple modulations occur as textures with localized
twists in two or more spatial directions \cite{Hornreich82,Wright89}. 
In noncentrosymmetric magnetic systems 
with Lifshitz invariants these double-twist motifs 
exist as smooth (baby)-Skyrmions \cite{JETP89,JMMM94,Nature06}, 
static solitonic textures localized in two spatial directions,
which can be extended into the third direction as Skyrmion strings or Hopfions.
These magnetic Skyrmions are stabilized solely 
by the chiral DM couplings \cite{Nature06,JMMM94}, 
which prevent a spontaneous collapse 
into topological singularities. 
%
\textit {Skyrmionic matter} is created by the condensation 
of these solitonic units, similar to 
vortex matter in type-II superconductors \cite{JETP89}.
Just such chiral Skyrmions have been recently observed
in thin layers of noncentrosymmetric ferromagnet
(Fe,Co)Si \cite{Yu10}.
Skyrmionic states stabilized by Lifshitz-type invariants
may exist and form extended mesophases in various 
condensed matter systems, as chiral liquid crystals, 
ferroelectrics, multiferroics, and in confined achiral systems
(e.g., thin magnetic layers) \cite{PRL01,Bode07,Wright89,Tkalec09}.

Skyrmionic textures usually form through nucleation, 
following a classification introduced by DeGennes \cite{DeGennes75} 
for (continuous) transitions
into incommensurate modulated phases. 
As demonstrated in Refs.~\cite{JETP89,JMMM94},
for the low-temperature micromagnetic model of chiral magnets,
at the transition from the field-driven Skyrmion lattices 
into the polarized homogeneous state the lattice period
diverges and the Skyrmions are set free as localized excitations.
As the Skyrmions retain their size and axisymmetric shape, 
there is a full spectrum of lattice modes up 
to the transition, in contrary to an instability type 
transition where  the amplitude of 
one fundamental mode acts 
as small parameter of the transition  \cite{DeGennes75}.

Here, we show for the standard model of chiral 
isotropic ferromagnets \cite{Dz64,Bak80,Nature06}
that Skyrmions are confined 
very close to the ordering temperature.
In that part of the phase diagram, the 
creation of Skyrmions as stable units and 
their condensation into extended textures occurs 
simultaneously through a rare case of an instability-type 
nucleation transition \cite{Felix86},
but the confined Skyrmions as discernible units 
may arrange in different mesophases.
This is a consequence of the coupling 
between the magnitude and the angular part 
of the order parameter.
Thus, near the ordering transitions, 
the local magnetization is not only multiply 
twisted but also longitudinally modulated.
From numerical investigations on 2D models of 
isotropic chiral ferromagnets, a staggered half-Skyrmion square lattice 
at zero and low fields and a hexagonal 
Skyrmion lattice at larger fields are found 
in overlapping regions of the phase diagram near the 
transition temperature.
Furthermore, the thermodynamic stability of Skyrmionic states 
can be favoured 
with respect to one-dimensional modulations 
by supplementing the model with cubic exchange anisotropy or in
a modified model for metallic chiral magnets \cite{Nature06}.

Within the standard phenomenological (Dzyaloshinskii) 
theory  \cite{Dz64} the magnetic energy density 
of a noncentrosymmetric ferromagnet 
can be written in the dimensionless form
\cite{Bak80,Nature06}
\begin{eqnarray}
\Phi 
  =(\mathbf{grad}\: \mathbf{m})^2
- w_D (\mathbf{m})
-h(\mathbf{n} \cdot \mathbf{m}) + a m^2+ m^4.
\label{density}
\end{eqnarray}
Here, reduced values of the spatial
variable $\mathbf{x} = \mathbf{r}/L_D$, 
the magnetization $\mathbf{m} = \mathbf{M}/M_0$, and
the applied magnetic field $h \mathbf{n}$ ($h = |\mathbf{H}|/H_0$,
$\mathbf{n}$ is a unity vector along $\mathbf{H}$),
are expressed via the parameters
of the energy density  \cite{Bak80}
$w(\mathbf{M}) =A (\mathbf{grad} \mathbf{M})^2
 + Dw_D (\mathbf{M})
 - \mathbf{H} \cdot \mathbf{M} + f(M)$
 (where $f(M)= a_1 M^2 + a_2 M^4$, and $ M = | \mathbf{M}|$):
$L_D = A/D$, $H_0 = \kappa M_0$, $M_0 = (\kappa/a_2)^{1/2}$,
$a = a_1/\kappa$, $\kappa = D^2/(A)$.
DM energy $w_D$ consists of Lifshitz invariants
\cite{Dz64}
\begin{eqnarray}
\mathcal{L}_{ij}^{(k)} = m_i(\partial m_j/\partial x_k)
-m_j(\partial m_i/\partial x_k).
\label{dm}
\end{eqnarray}
For noncentrosymmetric uniaxial ferromagnets
DM functionals $w_D$ are listed
in \cite{JETP89}.
In particular, for isotropic and cubic helimagnets
$w_D(\mathbf{m})=
\mathcal{L}_{xz}^{(y)}+\mathcal{L}_{zy}^{(x)}+\mathcal{L}_{xy}^{(z)}$
$= \mathbf{m}\cdot \mathrm{rot}\: \mathbf{m}$
\cite{Bak80} (for details see appendix).
Functional (\ref{density}) includes 
only the basic (\textit{isotropic})
interactions essential to stabilize 
chiral modulations.
This depends on three internal
variables (components of the magnetization vector
$\mathbf{m}$) and two control parameters,
the reduced magnetic field amplutude
$h$ and the  "effective" temperature $a(T)$.
In chiral magnetism models (\ref{density}) play
a fundamental role and are 
similar to the Frank energy in liquid
crystals and Landau-Ginzburg functional in 
superconductivity.
We consider 2D chiral modulations homogeneous
along the applied field $\mathbf{h} || z$
and modulated in the plane perpendicular
$\mathbf{h}$.

\begin{figure}
\includegraphics[width=8.25cm]{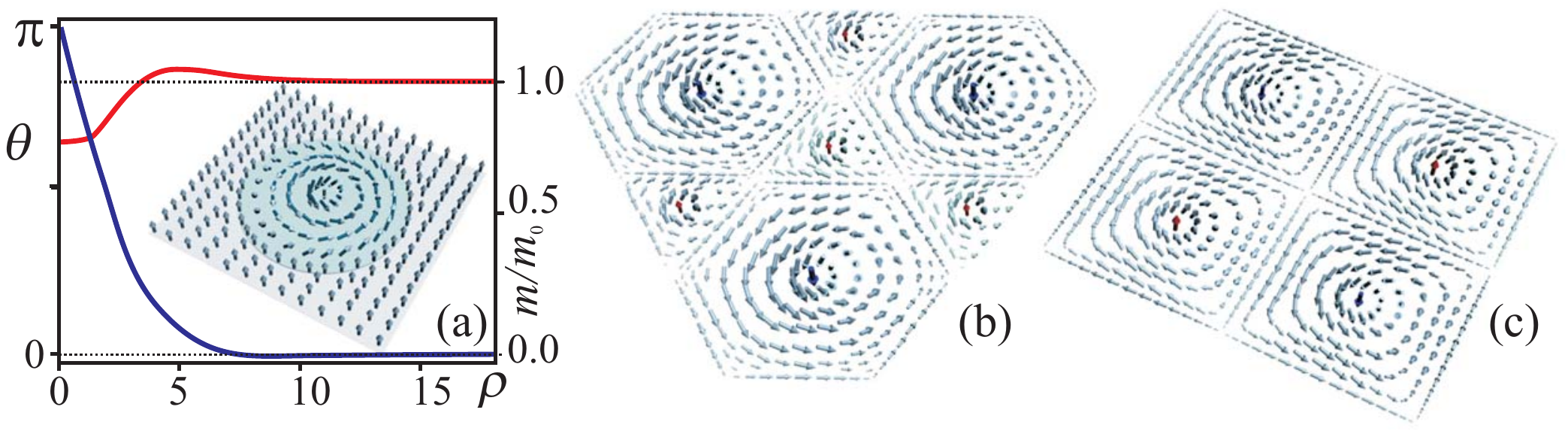}
\caption{
\label{f1a}
(Color online) 
(a) 
Equilibrium solutions of localized Skyrmions,
shown as cross section in the inset,
display strong localization of $\theta (\rho)$ (blue) 
and weak variation of the magnetization 
modulus $m (\rho)$ (red) in a broad range of temperature 
and field ($a, h$).
Exemplary profiles for $a=-0.25$, $h=0.25$.
%
(b) At lower fields, isolated Skyrmions condense 
into a dense-packed hexagonal lattice.
(c) Near the ordering temperature $a_c = 0.25$
square lattice solutions with 
staggered \textit{half-Skyrmion} cells arise.
}
\end{figure}

\textit{Isolated and embedded Skyrmions}. 
The equations minimizing functional (\ref{density})
include solutions for axisymmetric localized
states (\textit{isolated Skyrmions}), 
$\psi=\psi (\phi)$, and  $\theta(\rho)$, $m (\rho)$
where we use spherical coordinates for the magnetization
$(m,\theta,\psi)$  and cylindrical coordinates 
for the spatial variables $(\rho,\phi,z)$. 
The solutions $\psi(\phi)$ 
for all noncentrosymmetric 
classes have been derived in 
\cite{JETP89}. 
For cubic helimagnets and uniaxial
systems with $n$22 symmetry,
$\psi = \phi - \pi/2$ (Fig. \ref{f1a} a).
Profiles $\theta(\rho)$ and  $m (\rho)$
of isolated Skyrmions
are derived from numerical solutions of
the Euler equations, as given in Ref.~\onlinecite{Nature06} and common for all 
noncentrosymmetric classes with Lifshitz invariants,
%
with the boundary conditions
$\theta (0) =\pi$, $d m/d \rho (0) = 0$,
$\theta (\infty) = 0$, $ m (\infty) = m_0$
($m_0$ is the magnetization in the
saturated state).
%
For extended textures of two-dimensional models, 
the functional has been investigated 
by numerical energy minimization using finite-difference discretization 
on rectangular grids with adjustable grid spacings \cite{Nature06}.

\begin{figure}
\includegraphics[width=8.0cm]{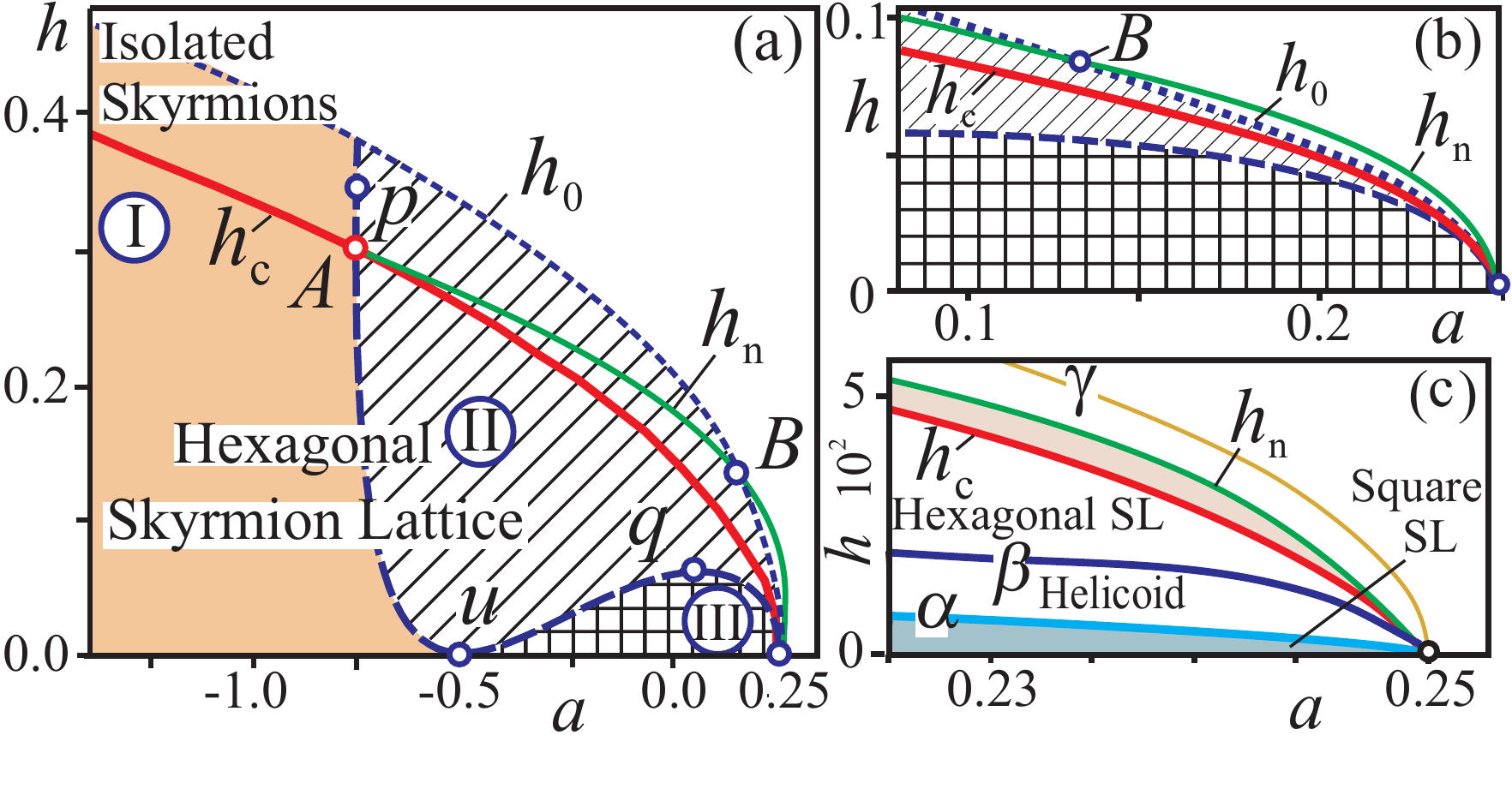}
\caption{
\label{f1b}
(Color online) 
Field $h$ vs. temperature $a$ phase diagram.
(a) Areas with
\textit{repulsive} (I) (shaded),
\textit{attractive} (II) Skyrmions (hatched),
and strictly \textit{confined} pocket (III)
(cross-hatched) are separated by $h^{\star}$ (dashed line), Eq.~(\ref{criticalline1}).
Isolated Skyrmions collapse at
critical line $h_0$. 
Above this line no static Skyrmions exist.
Below line $h_c$ 
Skyrmions condense into 
a hexagonal lattice
%
%
In region (II), the hexagonal Skyrmion lattice (SL)
exists as metastable state up to the nucleation field $h_n$.
For temperatures between points $A$ and $B$, 
$h_n < h_0$, for larger temperatures $a_B < a < a_c = 0.25$ 
the isolated Skyrmions disappear at lower fields 
than the dense SL, $h_0 > h_n$.
For clarity, line $h_n$ is only schematically 
given in panel (a), numerically exact data 
are shown in panel (b).
Detail of the phase diagram (c) near the ordering temperature
shows the existence regions
for different modulated
states.
Lines for first order transitions: 
$\alpha$  square half-SL $\leftrightarrow$ hexagonal SL,
$\beta$  helicoid $\leftrightarrow$ hexagonal SL.
Line $\gamma$ marks the continuous transition from
the conical equilibrium phase in isotropic systems 
to the paramagnetic phase.
}
\end{figure}

Isolated
Skyrmions (Fig.~\ref{f1a}~(a)) 
exist only below
a critical line $h_0$
and condense into a hexagonal lattice (Fig.~\ref{f1b})
below a field $h_c$,
which marks the transition between the Skyrmion lattice (SL)
and the homogeneous paramagnetic state (see Fig.~\ref{f1b}~(a), (b)).
Near the ordering temperature a square
\textit{half-Skyrmion} lattice (Fig.~\ref{f1a}~(c))
has lower energy than the hexagonal lattice.
Half-SLs consist of  cells with up and down magnetization
in the center and in-plane magnetization
along the cell boundaries. 
Such cells have a topological charge $1/2$.
In the hexagonal SLs
the magnetization at the
boundaries (center) is
(anti)parallel to the applied field. 
The cells
bear unit topological charges.

\textit{Confinement}.
By solving the linearized Euler equations
for  the asymptotics of isolated Skyrmions
$\Delta m =(m - m_0),\, \theta \propto \exp(-\kappa \rho)$
($\rho \gg 1$)
one finds three distinct regions in the magnetic
phase diagram with different character 
of Skyrmion-Skyrmion interactions (Fig.~\ref{f1b}~(a)):
\textit{repulsive} interactions 
in a broad temperature range (area (I))
are changed to \textit{attractive} 
interaction at higher temperatures (area (II)).
Finally in area (III)
near the ordering temperature $a_c=0.25$
strictly confined Skyrmions exist.
These regions are separated by the line
\begin{eqnarray}
h^{\star} =\sqrt{2 \pm  P(a)}(a+1 \pm P(a)/2), \;
P(a) = \sqrt{3+4a}\,
\label{criticalline1}
\end{eqnarray}
with turning points $p$ ($-0.75,\sqrt{2}/4$),
$q$ ($0.06,0.032\sqrt{5}$), and 
$u$ (-0.5, 0)
(dashed line in Fig.~\ref{f1b}~(a)).
In the major part of the phase diagram,
the evolution of SLs under a magnetic
field closely agrees with the behavior
studied earlier for the low-temperature limit \cite{JETP89,JMMM94}
and the transition mechanism at the high-field limit 
is of the nucleation type with isolated Skyrmion excitations 
appearing below the instability line $h_0$.
The temperature of the turning point 
$p$ ($a_p = -0.75$), a \textit{confinement}
temperature $T_c$ determines
the frontier between low- ($T < T_p$) 
and high-temperature ($T_p < a < T_c$) chiral
modulations. 
Usually $\Delta T = T_c - T_p \ll T_c$
and high-temperature modulations
are restricted by a narrow vicinity
of the Neel temperature.
According to calculations
in \cite{Nature06} for MnSi high-temperature
Skyrmions exist in temperature interval
$\Delta T = T_c - T_p$ = 2 K below the
Curie temperature $T_c$.

\begin{figure}
\includegraphics[width=8.0cm]{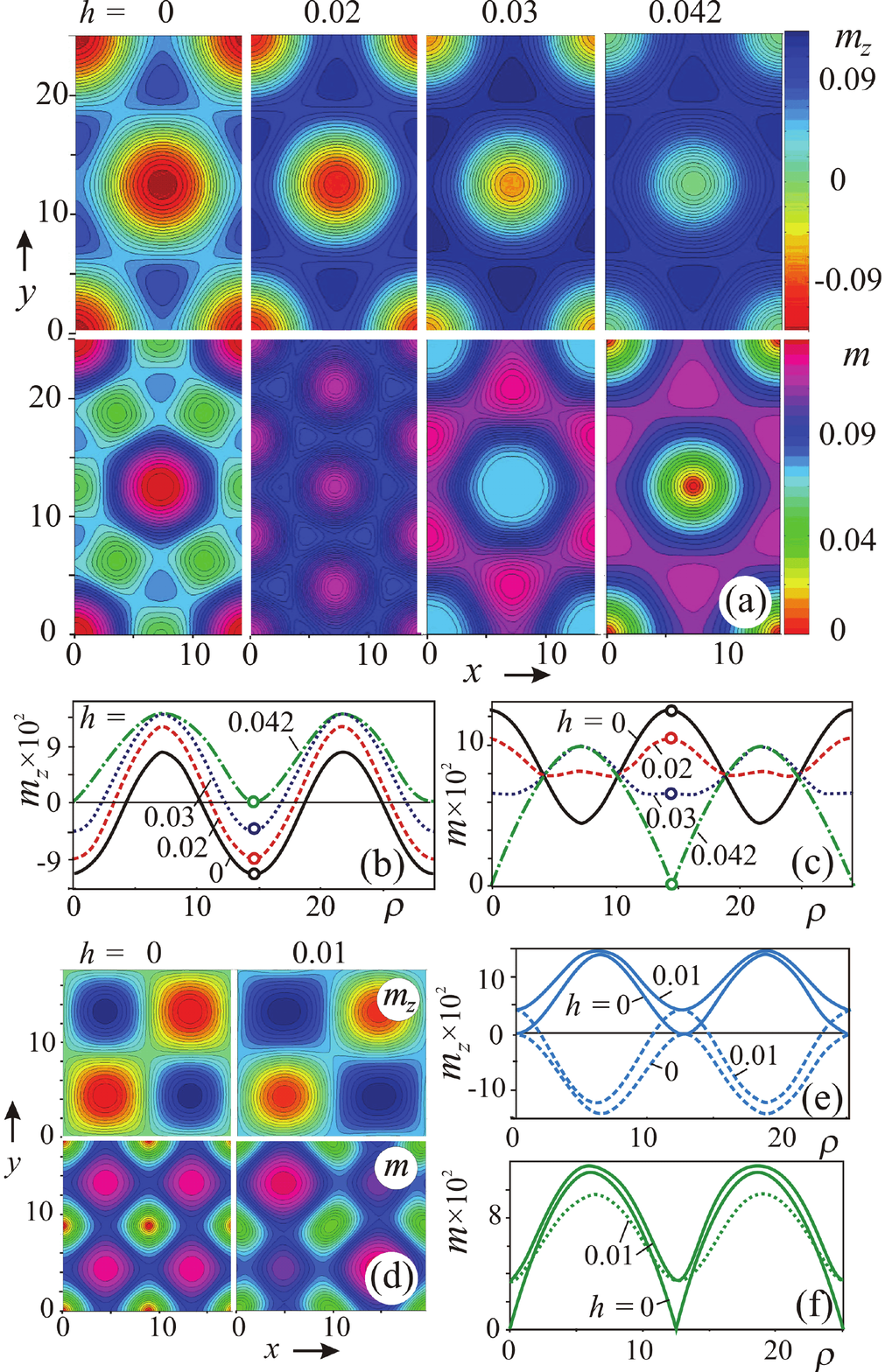}
\caption{%
\label{f2}
(Color online) 
Numerically exact solutions
for hexagonal (a)--(c)
and square (d)--(f) Skyrmion 
lattices within the
\textit{confinement} pocket.
Contour plots  of $M_z (x,y)$ and
$M (x,y)$ (a),(d) and their
diagonal cross-sections (b),( e), (c),(f) 
are derived by minimization of  $\Phi$ (\ref{density}) 
with $a =0.23$ and different
values of the applied field.
%
Solid (dashed) lines in panels (e), (f) 
indicate profiles for cells 
with (anti)parallel magnetization 
in the center.
} 
\end{figure}

\textit{Evolution of confined Skyrmions}.
In the vicinity of the ordering
temperature the Skyrmion lattices 
exist within the confinement
pocket (III). In this region
Skyrmion states drastically differ
from those in the main part of the
phase diagram. 
Due to the "softness" of the magnetization
modulus the field-driven transformation of
the Skyrmion lattices evolves 
by distortions of the modulus profiles $m(\rho)$ 
both in the hexagonal 
and square  Skyrmion
lattices while the equilibrium
periods of the lattices do
not change strongly with increasing applied field (Fig. \ref{f2}). 
Despite the strong transformation of their internal structures 
the Skyrmion lattices preserve 
\textit{axisymmetric} distribution of the
magnetization near the centers of the Skyrmion lattice cells
(Fig. \ref{f2}).
This remarkable property reflects 
the basic physical mechanism
underlying the formation of Skyrmion
lattices.
The local energetic advantage 
of Skyrmion lattices over helicoids is
due to a larger energy reduction in the 
``double-twisted'' Skyrmion cell core 
compared to ``single-twisted'' helical states
\cite{Nature06,Wright89}.
This explains the unusual axial symmetry
of the cell cores and their stability.
An increasing magnetic field gradually
suppresses the antiparallel magnetization
in the cell core reducing the energetic advantage of
the ``double-twist'' and increases the overall energy 
of the condensed Skyrmion lattice.
At critical field $h_c$ a first-order transition
occurs into the polarized paramagnetic state. 
In the interval $h_c < h < h_n$ the hexagonal Skyrmion lattice
exists as a metastable state.  
At the lability field 
$h_n$ the magnetization modulus in the cell center
becomes zero (see magnetization profile 
for $h = 0.042$ in Fig.~\ref{f2} (c)). 
At this field, the "double-twist" 
region is suppressed, 
and the lattice loses its stability. 

At zero field with increasing temperature $a$ 
the magnetization modulus $m$ in 
hexagonal and square lattices gradually
decreases to zero at the ordering temperature.
This is the instability-type nucleation transition into 
the paramagnetic phase :
the order parameter $m$ becomes zero at 
the transition point (as in the instability mode), however,
the lattices retain their symmetry and the arrangement of axisymmetric 
Skyrmions up to the critical point.
In finite fields in the region (II) 
the attractive Skyrmion-Skyrmion interaction 
means that multi-Skyrmions can always form bound states.
Near $a_c$ such clusters of Skyrmions 
are more stable than the isolated Skyrmions.
This is seen in Fig.~\ref{f1b}~(a),(b). 
For temperatures above point $B$, $a_B < a < a_c$, 
the metastable SL as the densest infinite cluster 
is more stable than isolated Skyrmions.
Thus, in region (II) of the phase diagram, 
there is clustering and, for higher temperatures,
confinement of isolated Skyrmion excitations.

The predicted Skyrmionic textures
and the pre-cursor phenomena associated with 
the confinement of Skyrmions near the ordering 
transition are observable in magnetic materials
with appropriate symmetry.
In a large group of {\textit{uniaxial noncentrosymmetric magnets}}
the DM energy is described by gradients 
only along directions perpendicular to
the axis (e.g. multiferroic BiFeO$_3$, space group $R3c$
and antiferromagnetic Ba$_2$CuGe$_2$O$_7$, space group $P \bar{4}2_1m$ 
\cite{PRB02,Sosnowska82}).
In  such magnets one-dimensional
modulations with the propagation vector
in the basal plane (\textit{helicoids})
exist in broad ranges of the magnetic
fields \cite{Dz64,Mukamel85}.
In the confinement pocket of the phase diagram (III) 
the helicoids also exist only as bound
states \cite{Mukamel85}.
They have the lowest energy at lower
fields, and the hexagonal SL corresponds to the global 
energy minimum at higher fields (Fig.~\ref{f1b}~(c)).

\begin{figure}
\includegraphics[width=8.0cm]{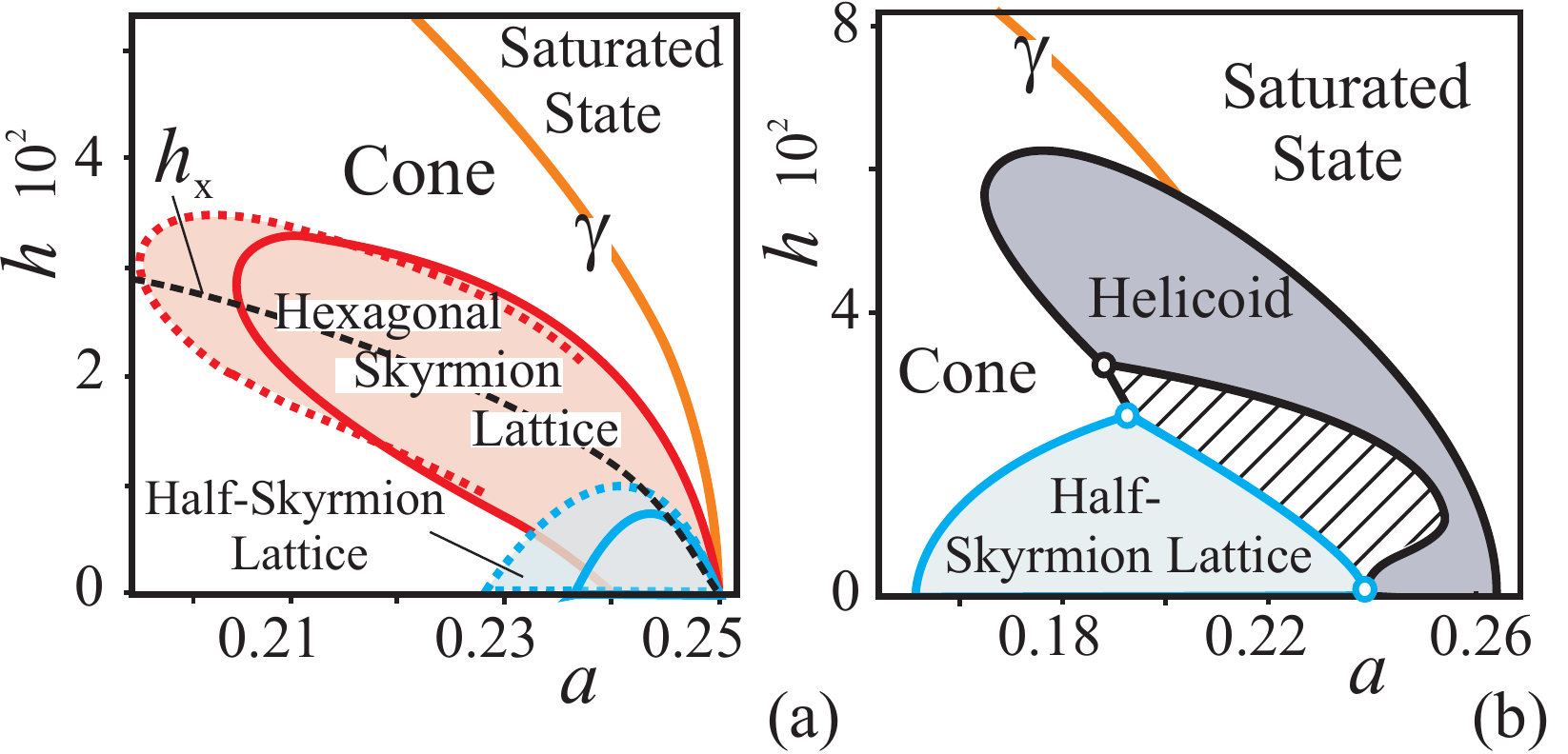}
\caption{
\label{f4}
(Color online) 
Magnetic phase diagrams 
of cubic helimagnets
with exchange anisotropy
$b  = - 0.05$
and the applied field
along (111) (solid)
and (001) (dashed)
axes (a) contains
regions  with thermodynamically
stable hexagonal SLs and half-SLs.
Magnetic phase diagram
of the modified isotropic model
with  $\eta =0.8$ (b)
has extended domains with
thermodynamically stable
helicoid, half-SLs and hexagonal SLs 
with the magnetization along
the cell axes parallel
to the magnetic field
(hatched area).
}
\end{figure}

\textit{Cubic helimagnets.} In other groups of chiral
magnets the DM energy includes contributions
with gradients along all three spatial directions. 
%
%
This stabilizes chiral modulations with propagation
along the direction of an applied field as \textit{cone} phases \cite{Bak80}.
For the isotropic model $\Phi(\mathbf{m})$
(\ref{density}) the cone phase solution 
with the fixed magnetization modulus 
and rotation of $\mathbf{m}$
around the applied magnetic field:
$\psi = z,\; \cos (\theta) = h/m, \;  m = |a-1/2|/2$,
is the global energy minimum in the whole region
where modulated states exist.
For cubic helimagnets, the energy density (\ref{density}) 
has to be supplemented by anisotropic contributions,
$\Phi_a = b \sum_{i}(\partial m_i/\partial x_i)^2 +k \sum_{i}m_i^4 $,
where $b$ and $k$ are reduced values of exchange
and cubic anisotropies \cite{Bak80}.
These anisotropic interactions
impair the ideal harmonic twisting of the cone phase 
and lead to the thermodynamic stability of Skyrmion states, as shown in the 
equilibrium phase diagram Fig.~\ref{f4}~(a).
The difference 
between the energy of the hexagonal Skyrmion lattice $W_{sk}$
and of the cone phase $W_{cone}$ calculated for the isotropic model,
$\Delta W_{min}= W_{sk}- W_{cone}$,
has minima along a curve $h_{\mathrm{x}}(a)$
which reaches the critical point $h_{\mathrm{x}}(a_c)=0$
as $\Delta W_{min} = 0.0784 (0.25-a)$).
Weak exchange anisotropy of a cubic helimagnet, therefore, 
creates a pocket around $a_c$, where the hexagonal 
Skyrmion lattice becomes the global
energy minimum in a field (Fig. \ref{f4} (a)).
This case is realized in cubic helimagnets
with negative exchange anisotropy ($b < 0$)
as in MnSi \cite{Bak80}. 
This anisotropy effect provides a basic mechanism,
by which a Skyrmionic texture is stabilized 
in applied fields, as observed experimentally 
as so-called "A-phase" in MnSi \cite{Gregory92,Lebech95,Muhlbauer09}.
The exchange anisotropy $b < 0$ also leads 
to the thermodynamic stability of half-SLs (Fig.~\ref{f4}~(a)). 
The stabilization of such textures 
may be responsible for anomalous
precursor effects in cubic helimagnets 
in zero field \cite{Pfleiderer04,Nature06,Pappas09}.

\textit{Chiral modulations in non-Heisenberg models}.
A generalization of isotropic chiral magnets 
proposed in \cite{Nature06} replaces 
the usual Heisenberg-like exchange model 
by a non-linear sigma-model coupled to a modulus 
field with different stiffnesses.
This yields a generalized gradient energy for a chiral isotropic 
system with a vector order parameter, which is equivalent to the 
phenomenological theory in the director formalism \cite{Wright89,Nature06}
$  \sum_{i,j} (\partial_i m_j)^2 \rightarrow
 \sum_{i,j} (\partial_i m_j)^2
 + (1 -\eta) \sum_{i,j} (\partial_i m)^2$
=  
$ m^2 \sum_{i,j} (\partial_i n_j)^2 + \eta \sum_i(\partial_i m)^2$.
Parameter $\eta$ equals unity 
for a ``Heisenberg'' model, in chiral nematics $\eta = 1/3$ \cite{Wright89}. 
%
%
Confined chiral modulations are very sensitive to values of $\eta < 1$.
The magnetic phase diagram calculated for 
$\eta =0.8$ includes pockets with square half-SLs, hexagonal SLs with
the magnetization in the center of the cells
parallel to the applied field,
and helicoids with propagation transverse 
to the applied field (Fig. \ref{f4} b).

%
The confinement effects on chiral Skyrmions strongly changes
the picture of the formation and evolution 
of chiral modulated textures 
and shed new light on the problem of precursor states
observed as blue phases in chiral nematics \cite{Wright89} 
and in chiral magnets \cite{Gregory92,Lebech95,Pfleiderer04,Nature06,Pappas09,Muhlbauer09}.
The results show that confinement / deconfinement 
transitions of localized string-like solitons
can be realized in these condensed matter systems.
They provide counterparts of formation mechanisms 
for extended microscopic matter from topological solitons,
as devised originally in the Skyrme model \cite{Skyrme61}. 

\textit{Acknowledgment.} 
We thank 
K. v. Bergmann, G. Bihlmayer, S. Bl\"{u}gel, H. Eschrig, S.V. Grigoriev, 
J.-H. Han, S. Heinze, R. M\"{o}ssner, C. Pappas, W. Selke, H. Wilhelm, 
and M. Zhitomirsky for discussions. 
Support by DFG project RO 2238/9-1 is gratefully acknowledged.


\begin{appendices}

\section{APPENDIX:Dzyaloshinskii theory of chiral helimagnetism}

The appendix includes technical 
comments on the mathematical structure of the interaction functional
(\ref{density}),
on the mathematical relation between this model 
and those for different classes of noncentrosymmetric
magnetic systems and chiral liquid crystals.
The equations for isolated
and bound Skyrmions are provided
along with a short information on 
the numerical methods to solve these problems.
%


\subsection{1. Basic model.}
In noncentrosymmetric ferromagnets chiral asymmetry
of exchange coupling related to the quantum-mechanical
Dzyaloshinskii-Moriya interactions \cite{Dz57}
induces long-period modulations of the magnetization
vector $\mathbf{M}$ with a fixed rotation sense \cite{Dz64}.
The equilibrium chiral modulations are derived by
minimization of the magnetic energy which can be
written in the following general form \cite{Dz64}
\begin{equation}
W= A \sum_{i,j}(\partial M_j/\partial x_i)^2
-\mathbf{M}\cdot\mathbf{H}+ W_0 (\mathbf{M})+W_D (\mathbf{M})
\label{density5}
\end{equation}
where  $A$ is the exchange stiffness constant,
 $\mathbf{H}$ is an applied magnetic
field, $W_0$ collects short-range magnetic interactions
independent on spatial derivatives of the magnetization, and
Dzyaloshinsky-Moriya energy $W_D$  
is composed of antisymmetric terms
linear with respect to first spatial derivatives
of the magnetization vector $\mathbf{M} (\mathbf{r})$ 
$$\mathcal{L}_{ij}^{(k)} = M_i \left(\partial M_j/\partial x_k \right)
-M_i \left(\partial M_j/\partial x_k \right).$$
Functional forms of $W_D$ energy contributions are determined by
crystallographic symmetry of a noncentrosymmetric magnetic crystal
\cite{Dz64,Bak80,JETP89}. 
Particularly, for important cases of noncentrosymmetric ferromagnets
belonging to cubic (23, 432) and uniaxial crystallographic
classes ($n$mm, $\bar{4}$2m) 
the DM energy contributions have the following
form \cite{Dz64,Bak80,JETP89}
\begin{eqnarray}
&& W_D=D\,(\mathcal{L}_{yx}^{(z)}+\mathcal{L}_{xz}^{(y)}+\mathcal{L}_{zy}^{(x)})
=D\,\mathbf{M}\cdot \mathrm{rot}\mathbf{M} \nonumber \\
&& \quad \quad \quad \quad \quad \quad \quad \quad \quad \quad \, \, \, \,\mathrm{for} \; (23), (432);
\nonumber \\
&& W_D=D\,(\mathcal{L}_{xz}^{(x)}+\mathcal{L}_{yz}^{(y)}) \quad \mathrm{for} \; (nmm); 
\quad \quad
\nonumber \\
&& W_D=D\,(\mathcal{L}_{xz}^{(y)}+\mathcal{L}_{yz}^{(x)}) \quad \mathrm{for} \; (\bar{4}2m) 
\label{LifshitzU}
\end{eqnarray}
where $n = 3, 4, 6$ and $D$ is
a Dzyaloshinskii constant. 
For other noncentrosymmetric classes
this $W_D$ energy includes two or more Dzyaloshinskii constants
\cite{JETP89}.
Near the ordering temperatures the magnetization
amplitude varies under the influence of the applied
field and temperature. 
Commonly this process is described by
including into the magnetic energy an additional
term \cite{Bak80}
\begin{eqnarray}
W \rightarrow W +  a_1 M^2 + a_2 M^4, \,
 a_1 = J(T-T_c), \,  a_2 > 0
\label{densitya}
\end{eqnarray}

This model is fundamental to the systematic phenomenological 
description of magnetic states in noncentrosymmetric magnetic
systems \cite{Dz64}. Interaction functional (\ref{density5})
plays in chiral magnetism a similar role as
the \textit{Frank energy} in liquid crystals \cite{DeGennes}
or \textit{Ginzburg-Landau} functional in physics
of superconductivity \cite{Brandt95}.
Its form is dictated by the natural
extension of Landau's approach to ordering transtions for
modulated systems which break the third Lifshitz condition, 
as pioneered by Dzyaloshinskii in particular 
in magnetism \cite{Toledano87}.
As a result, this type of theory constitutes 
the canonical form of statistical field theories 
for condensed matter systems, where particular couplings
and symmetry enable Lifshitz invariants to induce 
modulated phases.
Extensions including higher-order secondary effects
in magnetism (like anisotropies or dipolar interactions)
are widely used to describe evolution of modulated states 
in many chiral magnets \cite{Izyumov84,PRB02,JMMM94,Nature06}, 
including metamagnetic transitions 
in cubic helimagnets induced by high pressure 
\cite{Bak80,MnSi,Plumer81}.

By rescaling the spatial variable in 
(\ref{density}),(\ref{densitya})
$\mathbf{x} = \mathbf{r}/L_D$,
the magnetic field  $\mathbf{h} = \mathbf{H}/H_0$,
the magnetization $\mathbf{m} = \mathbf{M}/M_0$
\begin{eqnarray}
&&L_D = A/D, \quad 
H_0 = \kappa M_0, \;
M_0 = (\kappa/a_2)^{1/2},\nonumber \\ \;
 &&\kappa = D^2/(2A), \;
a = a_1/\kappa = J(T-T_c)/\kappa.
\label{units1}
\end{eqnarray}
energy $W$ (\ref{densitya}) can be written in
the following reduced form
(Eq. (\ref{density}))
\begin{eqnarray}
\Phi 
  =(\mathbf{grad}\: \mathbf{m})^2
- w_D (\mathbf{m})
-h(\mathbf{n} \cdot \mathbf{m}) + a m^2+ m^4,
\label{density2}
\end{eqnarray}
where $h = |\mathbf{h}|$ and $\mathbf{n}$
is a unity vector along the applied magnetic
field.
Functional (\ref{density2}) includes three internal
variables (components of the magnetization vector
$\mathbf{m}$) and two control parameters,
the reduced magnetic field amplutude
$h$ and the  "effective" temperature $a(T)$
(\ref{units1}).
By direct minimization of Eq. (\ref{density2})
we have derived one-dimensional (kinks, helicoids,
conical helices or spirals) and two-dimensional solutions
(isolated and bound Skyrmions) for arbitrary values
of the control parameters. 
These results are collected  in the phase diagram 
of solutions (Fig. 2). 

Energy (\ref{density2}) includes only basic interactions
\textit{essensial} to stabilize Skyrmionic and helical
phases. Solutions for chiral modulated phases 
and their most  general features attributed
to all chiral ferromagnets are determined
by interactions functional (\ref{density2}).
Generically, there are only small energy differences 
between these different modulated states.
On the other hand, weaker energy contributions (as 
magnetic anisotropy, stray-fields, magneto-elastic couplings) 
impose distortions on solutions of model (\ref{density2})
which reflect crystallographic symmetry and 
values of magnetic interactions in individual
chiral magnets.
These weaker interactions determine the stability 
limits of the different modulated states (Fig. 4). 
The fact that thermodynamical stability of  individual phases
and conditions of phase transfomations between them
are determined by magnetocrystalline anisotropy
and other relativistic or weaker interactions  means 
that (i) the basic theory only determines a set 
of different and unusual modulated phases, while
(ii) the transitions between these modulated states, 
and their evolution in magnetization processes depends
on symmetry and details of magnetic secondary effects in chiral magnets,
in particular the strengths of relativistic magnetic interactions.
Thus functional (\ref{density2}) is  
the \textit{generic} model for a manifold
of interaction functionals describing
different groups of noncentrosymmetric magnetic crystals,
because it allows to identify the basic modulated
structures that may be found in them.

\subsection{2. Equations for isolated and bound Skyrmions}

The Euler equations for functional (\ref{density2})
have solutions for axisymmetric isolated structures
of type $\psi = \psi (\phi)$, $\theta(\rho)$, 
$m (\rho)$ (we introduce here spherical coordinates 
for the magnetization $(m,\theta,\psi)$  
and cylindrical coordinates for the spatial variables $(\rho,\phi,z)$)
\cite{Nature06}.
Solutions $\psi = \psi (\phi)$ are known 
for all uniaxial and cubic noncentrosymmetric
ferromagnets \cite{JETP89}. 
Particularly,
$\psi=\phi-\pi/2$ for cubic helimagnets,
$\psi=\phi$  and $\psi=-\phi-\pi/2$ 
for uniaxial ferromagnets with
$(nmm)$ and $\bar{4}2m$ symmetry, correspondingly.
After substitution of solutions $\psi = \psi (\phi)$
into (\ref{density2}) and integration with respect to
$\phi$ the total energy  $E$ of a Skyrmion
(per unit length along $z$) 
is $E = 2\pi  \int_0^{\infty} \Phi (m, \theta) \rho  d \rho$
with energy density

\begin{eqnarray}
&&\Phi = m_{\rho}^2+ m^2 \left[ \theta_{\rho}^2 +\frac{\sin^2 \theta}{\rho^2} 
- \theta_{\rho} - \frac{\sin \theta \cos \theta}{\rho} \right] 
+ a m^2+\nonumber\\
&&\quad \quad \quad \quad \quad \quad \quad+ m^4 - h m \cos \theta
\label{density3}
\end{eqnarray}
where a common convention $ \partial f/ \partial x \equiv f_x$ is applied.
The Euler equations for the functional (\ref{density3}) 

\begin{eqnarray}
&& m \left[ \theta_{\rho \rho} +  \frac{\theta_{\rho}}{\rho}
-\frac{\sin \theta \cos \theta}{\rho^2} 
 - \frac{2\sin ^2 \theta}{\rho} - h \sin (\theta) \right]+\nonumber\\
 &&\quad \quad \quad+ 2 \left( \theta_{\rho} -1 \right) m_{\rho} = 0,
\nonumber \\
&& m_{\rho \rho }+ \frac{m_{\rho }}{\rho}
+m \left[ \theta_{\rho}^2 +\frac{\sin^2 \theta}{\rho^2} 
- \theta_{\rho} - \frac{\sin \theta \cos \theta}{\rho} \right]+\nonumber\\ 
&&\quad \quad \quad+ 2a m+ 4m^3 - h \cos (\theta) =0
\label{Euler1}
\end{eqnarray}
with boundary conditions $\theta(0) = \pi$, $\theta(\infty) = 0$,
$m(\infty) = m_0 $ describe the structure of isolated Skyrmions
(the magnetization of the homogeneous 
phase $m_0$ is derived from equation  $2a m+ 4m^3 - h =0$).
Typical solutions  $\theta (\rho)$, $m (\rho)$ of Eqs. (\ref{Euler1})
are plotted in Fig. 1.

For present work, we have evaluated numerically the solutions 
for the energy functionals of type Eq.~(\ref{density}) within a
standard approach using finite differences for gradient terms
and adjustable grids to accommmodate modulated states with periodic
boundary conditions. Search for modulated states and energy minimization 
was done by Monte Carlo simulated annealing. For 2-dimensional models, this 
approach can give  converged results, as checked by comparison 
with analytical results. 
Including secondary effects, the approach is able and has been 
used by us to calculate thermodynamic stability of phases, and to 
study the complete temperature-field phase diagrams in terms of
equilbrium (mean-field) states of the energy functionals Eq.~(\ref{density}).
As we are investigating here long-period modulated phases, the
phase diagrams are expected to give qualitatively the correct 
results for these phenomenological models.
\end{appendices}


%
%
%

\end{document}